\documentstyle[twoside,fleqn,espcrc2,epsf,epsfig]{article}

\def\simge{\mathrel{%
   \rlap{\raise 0.511ex \hbox{$>$}}{\lower 0.511ex \hbox{$\sim$}}}}
\def\simle{\mathrel{
   \rlap{\raise 0.511ex \hbox{$<$}}{\lower 0.511ex \hbox{$\sim$}}}}

\def\slashchar#1{\setbox0=\hbox{$#1$}           % set a box for #1
   \dimen0=\wd0                                 % and get its size
   \setbox1=\hbox{/} \dimen1=\wd1               % get size of /
   \ifdim\dimen0>\dimen1                        % #1 is bigger
      \rlap{\hbox to \dimen0{\hfil/\hfil}}      % so center / in box
      #1                                        % and print #1
   \else                                        % / is bigger
      \rlap{\hbox to \dimen1{\hfil$#1$\hfil}}   % so center #1
      /                                         % and print /
   \fi}                                         %
\newcommand{\ewxy}[2]{\setlength{\epsfxsize}{#2}\epsfbox[45 240 320 350]{#1}}
\newcommand{\be}{\begin{equation}}
\newcommand{\ee}{\end{equation}}
\newcommand{\bea}{\begin{eqnarray}}
\newcommand{\eea}{\end{eqnarray}}
\newcommand{\msb}{\overline{\rm{MS}}}
\newcommand{\mev}{\,{\rm MeV}}   
\newcommand{\gev}{\,{\rm GeV}}   
\newcommand{\ml}{\overline m _l}
\newcommand{\mup}{\overline m _u}
\newcommand{\md}{\overline m _d}
\newcommand{\ms}{\overline m _s}

\title{Light quark masses and CKM matrix elements from lattice QCD}

\author{V. Lubicz\address{Dip. di Fisica, Universit\`a di Roma Tre and 
        INFN, Sezione di Roma Tre, \\ Via della Vasca Navale 84, 
	I-00146 Roma, Italy}}

\begin{document}

\begin{abstract}
I give a brief overview of recent results from lattice QCD calculations
which are relevant for the phenomenology of the Standard Model. I discuss,
in particular, the lattice determination of light quark masses and the
calculation of those hadronic quantities, such as semileptonic form factors, 
decay constants and $B$-parameters, which are of particular interest for the 
analysis of the CKM mixing matrix and the origin of CP violation. 
\end{abstract}

\maketitle

\section{LIGHT QUARK MASSES}
Calculations of light quark masses are becoming, at present, one of the most 
important subject of investigation for lattice QCD. Indeed, in spite of their 
relevant phenomenological interest, light quark masses are still among the 
less known fundamental parameters of the Standard Model. 

In the study of quark masses, the lattice method provides a unique approach: 
quark masses, defined as effective couplings renormalized at short distances, 
can be in fact determined on the lattice from non-perturbative calculations
of hadronic quantities. An inspection of recent lattice results for quark
masses suggests, however, that in such calculations a better understanding and 
quantification of the systematic errors is still an important requirement. To 
be specific, let us consider the lattice determinations of the strange quark 
mass, $\ms$. A compilation of results for this mass, in the $\msb$ scheme, at 
the renormalization scale $\mu=2 \gev$, is shown in Table \ref{tab:mslatt}. 
All the results have been obtained in the quenched approximation.
%___________________________________________________________________________
\begin{table}[hbt]
\caption[]{\it Lattice results for the strange quark mass $\ms (2\gev)$ in 
the $\msb$ scheme. We denote by ${\cal O}(a^2)$ and NPR the results obtained 
by using the non-perturbatively ${\cal O}(a)$-improved action and 
non-perturbative renormalization respectively.}
\label{tab:mslatt}
\begin{tabular}{llc}
\hline  & \multicolumn{2}{l}{Yr $\quad \ms(2 \gev)/\mev$}  \\  \hline
APE   \cite{mq_ape94}    & $94 \quad 128 \pm 18 $ &   \\
LANL  \cite{mq_lanl}     & $96 \quad 100 \pm 21 \pm 10 $ &   \\
FNAL  \cite{mq_fnal}     & $96 \quad 95 \pm 16 $ &   \\
APE   \cite{mq_ape96}    & $96 \quad 122 \pm 20 $ &   \\
SESAM \cite{mq_sesam}    & $97 \quad 166 \pm 15 $ &   \\
CP-PACS \cite{mq_cppacs} & $97 \quad 135 \pm  7 $ & (from $m_\phi$)  \\
                         & $97 \quad 111 \pm  4 $ & (from $m_K$)   \\
JLQCD \cite{mq_jlqcd}    & $97 \quad  97 \pm  9 $ &   \\
QCDSF \cite{mq_qcdsf}    & $97 \quad 112 \pm  5 $ &  ${\cal O}(a^2)$  \\
APETOV \cite{mq_apetov}  & $97 \quad 111 \pm 15 $ &  ${\cal O}(a^2)$  \\
APE   \cite{mq_ape98}    & $98 \quad 130 \pm 2 \pm 18 $  &  NPR  \\
APE   \cite{mq_noi}      & 
\multicolumn{2}{l}{$98 \quad 121 \pm 13 \qquad {\cal O}(a^2)$ + NPR}  \\
\hline
\end{tabular}
\vspace{-0.8truecm}
\end{table}
%___________________________________________________________________________
As can be seen from the table, although the lattice predictions all lie in the 
range $100 \simle \ms \simle 160 \mev$, however, within the errors quoted by 
the authors, the results obtained from different numerical studies are often 
in disagreement. 

The main two sources of systematic errors, which are responsible for the 
discrepancies discussed above, are easily identified. The quark mass that is 
directly computed in lattice simulations is the (short-distance) bare 
lattice quark mass $m(a)$, where $a$ is the lattice spacing. This quantity is 
typically determined by fixing, to its experimental value, the mass of a 
hadron containing a quark with the same flavour. Since the non-perturbative
lattice calculation of hadronic quantities is affected by discretization 
errors (in general of ${\cal O}(a)$), the same systematic errors will also 
propagate into the final determination of the quark masses.

The second step in these calculations consists in relating the bare quark mass
$m(a)$ to the renormalized mass $m(\mu)$, in a given renormalization scheme. 
The connection is provided by a multiplicative renormalization constant,
$m(\mu) = Z_m(a\mu) \, m(a)$. The perturbative, renormalization group improved 
expression of $Z_m(a\mu)$, at the NLO, has been given in ref.~\cite{mq_ape94}. 
Clearly, this matching introduces higher orders perturbative errors, which in 
turn affect the determination of the quark mass.

It is a remarkable fact that, at present, both discretization and higher 
orders perturbative errors, in the lattice determination of light quark masses, 
can be (and in fact have been) drastically reduced. The way to deal with 
discretization errors has been suggested by Symanzik \cite{sym} and 
consists in improving the lattice action and operators. This program has been
realized perturbatively in \cite{sw,heatlie} and, more recently, carried out 
at the non-perturbative level in %by the Alpha Collaboration 
\cite{alfa1}-\cite{alfa3}. The discretization errors are reduced to 
${\cal O}(a^2)$ in the chiral limit. The residual cutoff effects are expected 
to affect the determination of light quark masses by less than 1\% 
\cite{mq_noi}.

Higher orders perturbative errors in the evaluation of the quark mass 
renormalization constant are eliminated by adopting the non-perturbative 
renormalization prescription proposed in ref.~\cite{npr}. The idea consists 
on imposing the renormalization conditions directly on non-perturbatively 
calculated correlation functions between external off-shell quark states. This 
prescription defines the mass in a specific scheme, called RI-MOM scheme. 
Lattice calculations in this scheme are thus completely non-perturbative. 
Perturbation theory only enters if one wants to convert the result to the 
mass in the $\msb$ scheme. In this step (which is not necessary anyway) higher 
orders perturbative errors are expected to be negligible, since the relevant 
conversion factor is presently known at the N$^2$LO \cite{fl}.

Among the lattice predictions for the strange quark mass (see Table 
\ref{tab:mslatt}), the smallest values, corresponding to $\ms \simle 100 \mev$,
have been obtained in refs.~\cite{mq_lanl,mq_fnal,mq_jlqcd} as a result of an 
extrapolation to the continuum limit, $a \rightarrow 0$. However, the 
${\cal O}(a)$-improved calculations of refs.~\cite{mq_qcdsf,mq_apetov,mq_noi}, 
in which discretization effects are expected to be negligible, predict larger 
values, thus casting some doubts on the reliability of previous 
extrapolations.

\begin{figure}[htb]
\vspace{1.0truecm}
\hspace{-1.5truecm}
\ewxy{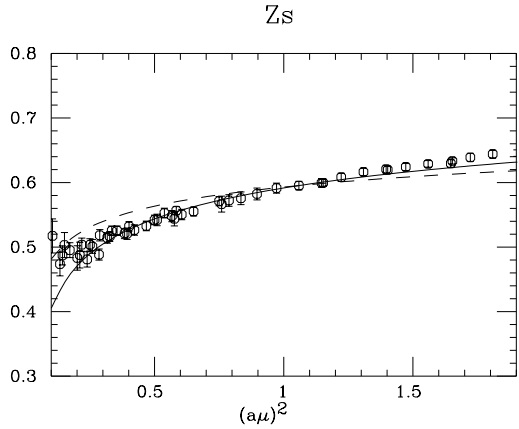}{60mm}
\vspace{0.7truecm}
\caption[]{\it $Z_S$, obtained by using the non-perturbative method, as a 
function of the renormalization scale. The dashed and solid curves represent 
the solutions of the renormalization group equations at the LO and NLO 
respectively.}
\protect\label{fig:zsfig}
\vspace{-0.5truecm}
\end{figure}

In ref.~\cite{mq_noi} the first lattice calculation of light quark masses, 
which combined both the use of a non-perturbatively improved action and the 
non-perturbative renormalization technique, has been presented. As already
noted in \cite{mq_ape98}, the non-perturbative calculation of renormalization 
constants removes the discrepancies, observed in previous calculations, 
between the results obtained by using different definitions of the bare quark 
mass (based on the vector and the axial-vector Ward identities respectively). 
It also gives values of the quark masses larger than those obtained by 
computing the renormalization constants in perturbation theory. The relevant 
renormalization constants which enter in the lattice calculations of the quark 
masses are those of the scalar and pseudoscalar densities, $Z_S$ and 
$Z_P$. By comparing the values of these constants obtained by using the 
non-perturbative method with the predictions of one-loop (boosted) 
perturbation theory it is found that the differences can be larger than 10\% 
and 30\% in the case of $Z_S$ and $Z_P$ respectively. Clearly, this error is
introduced in the determination of the quark masses, when the renormalization 
constants are evaluated in perturbation theory. As an example of a 
non-perturbative determination of renormalization constants, we show in 
Figure \ref{fig:zsfig} the values of $Z_S$ as a function of the 
renormalization scale $\mu$ \cite{mq_noi}. It is reassuring to find that 
this behaviour results in excellent agreement with the dependence predicted 
by the renormalization group equation.
%__________________________________________________________________________
%___________________________________________________________________________

As a final estimate of the average light ($\ml = (\mup + \md)/2$) and 
strange quark masses, in the $\msb$ scheme at the scale $\mu=2\gev$, I quote 
the results of ref.~\cite{mq_noi}:
\bea 
& \ml^{\scriptsize{NLO}} = 4.9(4) \, \mbox{MeV} \nonumber \\ 
& \ms^{\scriptsize{NLO}} = 121(13) \, \mbox{MeV}
\label{eq:mqnlo}
\eea 
and
\bea 
& \ml^{\scriptsize{N}^2\scriptsize{LO}} =  4.5(4) \, \mbox{MeV} \nonumber \\ 
& \ms^{\scriptsize{N}^2\scriptsize{LO}} = 111(12) \, \mbox{MeV}
\label{eq:mqn2lo}
\eea 
at the NLO and N$^2$LO respectively. In these estimates discretization and
higher orders perturbative errors are expected to be negligible, and the
remaining uncertainty is mainly due to the quenched approximation. The values 
in eq.~(\ref{eq:mqnlo}) are also in good agreement with recent results from 
QCD sum rules \cite{dominguez}.

\section{CKM MATRIX ELEMENTS}
In the analysis of the CKM mixing matrix, a particularly interesting issue is 
represented by the study of the unitarity triangle in the $\rho$-$\eta$ plane. 
Indeed, a non trivial shape of this triangle is the signature of a CKM source
of CP violation. 

\begin{figure}[htb]
\vspace{-1.7truecm}
\ewxy{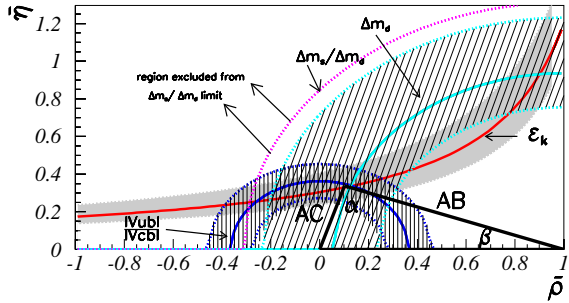}{45mm}
\vspace{3.0truecm}
\caption[]{\it The unitarity triangle in the
$\overline{\rho}$-$\overline{\eta}$ plane and the constraints derived from
$\Delta m_d$, $\Delta m_s/\Delta m_d$, $ \mid \epsilon_K \mid $ and 
$\left | V_{ub}\right | / \left | V_{cb}\right |$.}
\protect\label{fig:tria}
\vspace{-0.5truecm}
\end{figure}

At present, the phenomenological analysis of the unitarity triangle is 
constrained (mainly) by the existing measurements of $B^0_{d,s}-\bar 
B^0_{d,s}$  mixings, $K^0-\bar K^0$ mixing and semileptonic $b\rightarrow u$ 
transitions. Once these measurements are compared with the theoretical 
predictions, constraints on the $\rho$-$\eta$ parameters are derived. The 
status of the art is shown in Figure \ref{fig:tria}, from ref.~\cite{paganini}.
%___________________________________________________________________________
%___________________________________________________________________________
The theoretical predictions are based on the non-perturbative evaluation of 
some relevant hadronic matrix elements, which, in turn, are parameterized in 
terms of form factors, decay constants and $B$-parameters. The lattice method 
provide a reliable approach to perform such calculations. In the following, 
I will present a short compilation of these lattice results.

\subsection{$V_{ub}$ from $B \rightarrow \pi , \rho$ semileptonic decays}
\begin{figure}[htb]
\vspace{1.0truecm}
\hspace{0.4truecm}
\ewxy{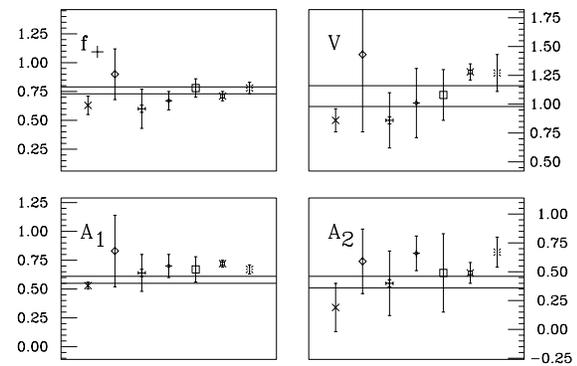}{50mm}
\vspace{1.0truecm}
\caption[]{\it Lattice results for the $D \rightarrow K , K^\ast \, l \nu$ 
form factors, at zero momentum transfer, \cite{lmms_sem}-\cite{wupp_sem}. 
The horizontal band indicates the present experimental average.}
\protect\label{fig:formf98}
\vspace{-0.5truecm}
\end{figure}

The determination of the $V_{ub}$ mixing angle is allowed by the analysis of 
$B \rightarrow \pi , \rho$ semileptonic decays. The reliability of lattice 
calculations in the study of heavy mesons semileptonic decays is provided by 
the analysis of $D$-meson decays, $D \rightarrow K , K^\ast \, l \nu$. In 
this processes the relevant mixing angle $V_{cs}$ is well constrained by the 
unitarity of the CKM matrix, so that the theoretical predictions can be 
compared with the experimental measurements. The comparison is shown in 
Figure \ref{fig:formf98}, where the lattice results for the four relevant 
form factors, at zero momentum transfer, are presented together with their 
experimental values \cite{ffexp}. A summary of these results is also given in 
Table \ref{tab:ffact}. 
%___________________________________________________________________________
%___________________________________________________________________________
%___________________________________________________________________________
\begin{table}[hbt]
\caption[]{\it Summary of lattice and experimental results for $D \rightarrow
K, K^\ast$ semileptonic decays form factors at zero momentum transfer. The 
results for $B \rightarrow \pi, \rho$ semileptonic decays, from 
ref.~\cite{ukb_sem}, are also presented.}
\label{tab:ffact}
\hspace{0.5truecm}
\begin{tabular}{lccc}
\hline  
& \multicolumn{2}{c}{$D \rightarrow K, K^\ast$} & $B \rightarrow \pi, \rho$ \\
\cline{2-3} \cline{4-4} 
& Lattice & Exp. & Lattice \\ \hline
$f_+(0)$  &  $0.72(7)$  &  $0.76(3)$  & $0.27(11)$ \\
$V(0)$    &  $1.14(18)$ &  $1.07(9)$  & $0.35(^{+6}_{-5})$ \\
$A_1(0)$  &  $0.64(8)$  &  $0.58(3)$  & $0.27(^{+5}_{-4})$ \\
$A_2(0)$  &  $0.53(13)$ &  $0.41(5)$  & $0.26(^{+5}_{-3})$ \\
\hline
\end{tabular}
\vspace{-0.5truecm}
\end{table}
%___________________________________________________________________________

For the $B \rightarrow \pi , \rho$ semileptonic decays, four different lattice 
groups have presented results so far \cite{elc_sem,ape_sem,wupp_sem,ukb_sem}. 
These are in quite good agreement one to each other. The more accurate 
determination of form factors has been obtained in ref.~\cite{ukb_sem}, from a 
combined analysis of $B$-meson semileptonic decays and radiative $B\rightarrow 
K^\ast \gamma$ decays. The results are shown in Table \ref{tab:ffact}.
 
\subsection{$\hat B_K$, the kaon bag parameter}
The bag parameter $\hat B_K$ is the relevant hadronic quantity which enters in 
the calculation of $\epsilon_K$. A recent compilation of lattice results gives,
for the NLO renormalization group invariant parameter, the estimate
\cite{gupta_bk}:
\be
\hat B_K = 0.86(6)(6)
\label{eq:bk_latt}
\ee
Notice that, in present analysis of the unitarity triangle, four constraints 
are used to determine the values of two parameters, $\rho$ and $\eta$ (see 
Figure \ref{fig:tria}). Thus, one of the constraints can also be removed, and
the corresponding quantity can be determined together with $\rho$ and $\eta$. 
In this way, in ref.~\cite{paganini} they obtain the value
\be
\hat B_K = 0.82 ^{+0.45}_{-0.24}
\ee
in very good agreement with the lattice determination 
of eq.~(\ref{eq:bk_latt}). This result also implies that significatively 
lower estimates of $\hat B_K$, such as those obtained by using the QCD 
hadronic duality approach ($\hat B_K=0.39 \pm 0.10$) \cite{pich} or using the 
$SU(3)$ symmetry and PCAC ($\hat B_K=1/3$) \cite{don} are presently strongly
disfavoured by the experiments.

\subsection{$\Delta m_d$ and $\Delta m_s$ from $B$-meson decay constants and 
$B$-parameters}
\begin{figure}[htb]
\vspace{0.5truecm}
\hspace{0.2truecm}
%\ewxy{fds.ps}{60mm}
\ewxy{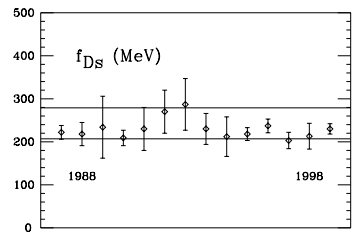}{40mm}
%\vspace{0.5truecm}
\caption[]{\it Lattice results for the pseudoscalar decay constant $f_{D_s}$. 
The horizontal band indicates the present experimental average. We refer to 
ref.~\cite{flysach} for a compilation of results and corresponding 
references.}
\protect\label{fig:fds}
\vspace{-0.5truecm}
\end{figure}

The theoretical estimates of the $B$-meson mass differences, $\Delta m_d$ and 
$\Delta m_s$, can be expressed in terms of the pseudoscalar decay constants, 
$f_{B_{d,s}}$, and $B$-parameters. A recent compilation of lattice results for 
these quantities has been presented in ref.~\cite{flysach} and it is shown
in Table \ref{tab:fBB}.
%___________________________________________________________________________
\begin{table}[t]
\caption[]{\it Summary of lattice results for $D$ and $B$-meson decay constant
and $B$ parameters}
\label{tab:fBB}
\begin{tabular}{ccc}
\hline
$f_D/(\mev)$ & $f_{D_s}/(\mev)$ & $f_{B}/(\mev)$ \\
200(30) & 220(30) & 170(35)  \\
\hline
$f_{B}\sqrt{B_B}/(\mev)$ & $f_{B_s}/f_{B}$ & $B_{B_s}/B_{B}$ \\
201(42) &  1.14(8) & 1.00(3)  \\
\hline
\end{tabular}
\vspace{-0.5truecm}
\end{table}
%___________________________________________________________________________
Notice that the lattice value for $f_{D_s}$, which has been predicted well 
before the first experimental measurement, it is in good agreement with the 
present experimental average, $f_{D_s}=243 \pm 36$ MeV~\cite{paganini}. The 
lattice results for this quantity have always been stable in the time. This is 
shown in Figure \ref{fig:fds}, where these results, obtained over a period of 
10 years, are presented together with the current experimental value.
%___________________________________________________________________________
%___________________________________________________________________________
The last point in the figure represents the preliminary result
of the first ${\cal O}(a)$-improved lattice calculation of the heavy mesons
decay constants. It corresponds to $f_{D_s}=(230 \pm 12)$ MeV~\cite{fbd_noi}.

The lattice determinations of the $B$-meson decay constant, $f_B$, still 
represent a genuine prediction, since this constant has not been measured yet 
in the experiments. From the (overconstrained) fit of the Standard Model, in
ref.~\cite{paganini} the value
\be
f_{B}\sqrt{B_B} = 213(21)(20) \mev
\ee
is obtained, in good agreement with the lattice determination (see Table 
\ref{tab:fBB}). An independent determination of $f_{B}$ can be also derived 
by combining the lattice determination of the ratio $f_{B}/f_{D_s}$, in which
many of the systematic errors are expected to cancel, with the experimental 
measurement of $f_{D_s}$. Two recent lattice calculations give:
\be
f_{B}/f_{D_s} = \left\{ 
\begin{array}{lr}
0.77(8) \qquad & {\rm APE~\cite{fbd_noi}} \\
0.75(5)(7) \qquad & {\rm MILC~\cite{fbd_milc}}
\end{array} 
\right. 
\ee
From these values I find:
\be
f_{B} = (185 \pm 27 \pm 19) \mev
\label{eq:fbrap}
\ee
where the first error comes from the experimental uncertainty on $f_{D_s}$
and the second one from the theoretical uncertainty on the ratio $f_{B}/
f_{D_s}$. Eq.~(\ref{eq:fbrap}) is also in good agreement with the direct 
determination of $f_{B}$ given in Table \ref{tab:fBB}.

\end{document}